# Enhanced Emission from WSe$_2$ Monolayers Coupled to Circular Bragg Gratings

Ngoc My Hanh Duong,[1] Zai-Quan Xu,[1,][*] Mehran Kianinia,[1] Rongbin Su,[2] Zhuojun Liu,[2] Sejeong Kim,[1] Carlo Bradac,[1] Lain-Jong Li,[3] Alexander Solntsev,[1] Jin Liu,[2,*] and Igor Aharonovich[1]

1. School of Mathematical and Physical Sciences, University of Technology Sydney, Ultimo, New South Wales, 2007, Australia
2. School of Physics, Sun Yat-sen University, Guangzhou 510275, China
3. Physical Science and Engineering (PSE) Division, King Abdullah University of Science and Technology (KAUST), Thuwal 23955-6900, Saudi Arabia

* zaiquan.xu@uts.edu.au, liujin23@mail.sysu.edu.cn

**Abstract**

Two-dimensional transition-metal dichalcogenides (TMDC) are of great interest for on-chip nanophotonics due to their unique optoelectronic properties. Here, we propose and realize coupling of tungsten diselenide (WSe$_2$) monolayers to circular Bragg grating structures to achieve enhanced emission. The interaction between WSe$_2$ and the resonant mode of the structure results in Purcell-enhanced emission, while the symmetric geometrical structure improves the directionality of the out-coupling stream of emitted photons. Furthermore, this hybrid structure produces a record high contrast of the spin valley readout (> 40%) revealed by the polarization resolved photoluminescence (PL) measurements. Our results are promising for on-chip integration of TMDC monolayers with optical resonators for nanophotonic circuits.

**Introduction**

Atomically-thin transition-metal dichalcogenides (TMDCs) are an emerging class of materials with unique optoelectronic properties.[1] They possess a direct bandgap[2], exhibit very large exciton binding energies[3] and strong spin-orbit coupling[4]. In addition, a very unique aspect of these materials, is a strong spin valley polarization (due to their natural inversion symmetry breaking), that allows optical readout[5-7] and electrical modulation even at room temperature.[8-

[11] These properties elevate TMDCs as one of the promising candidates for applications in nanophotonics, quantum optics, magneto-optics and valleytronics.[12-14]

Most of the promising applications of TMDCs—such as ultra-low-threshold lasers,[15] light emitting diodes (LEDs),[8,16] photodetectors[17] and spin valley readout rely on efficient generation/extraction (or absorption) of photons emitted by the TMDCs. Given the monolayer nature of the material, light absorption can be challenging; an emerging topic of research is the hybrid integration of TMDCs with photonic[15,18-19] or plasmonic cavities[20-21] to enhance light-matter interaction.

Circular Bragg grating (CBG) cavities, are structures which are rather simple to fabricate and produce high photon extraction rates and efficient Purcell enhancement.[22-23] The CBG is therefore particularly attractive and promising for integration with 2D materials. The basic structure consists of a wavelength-scale central disk surrounded by periodic concentric rings acting as an antenna. The circular grating operates under the second-order Bragg condition, and provides reflective feedback to the cavity and near-vertical upwards scattering. The emitted photons are firstly captured by the low-Q cavity formed by the central disk and the concentric circular grating, and are then scattered out of the plane by the circular grating with a convergent, far-field pattern for efficient photon collection by the objective.[24]

Here, we demonstrate efficient coupling between a monolayer $WSe_2$ and a CBG. A $WSe_2$ monolayer is transferred and spatially aligned with the photonic CBG cavity. When coupled, we observe a seven-fold enhancement in the excitonic emission from the hybrid $WSe_2$-CBG device. We also observe a Purcell enhancement of ~4, as deduced from measuring the time-resolved photoluminescence (PL) and high contrast of the spin valley readout.

A schematic illustration of the $WSe_2$ monolayer placed on top of the CBG is shown in figure 1a. $WSe_2$ flakes were grown by chemical vapour deposition (CVD).[25] The thickness of the flakes was measured to be ~0.7 nm using atomic force microscope (AFM) and confirming it to be a monolayer (figure. 1b).[17,26] The typical lateral length of the $WSe_2$ is ~50 μm. The monolayer was then transferred[26] onto a silicon nitride ($Si_3N_4$) substrate, which had been patterned with shallow CBG trenches. The large area of the $WSe_2$ allows for high precision in the positioning of the flake in the center of the CBG without requiring sophisticated alignment procedures. The emission from the flake is uniform across its entirety, which enables the easy and direct comparison of PL emission for the same WSe$_2$ monolayer from regions both in the center of the CBG and completely away from it.

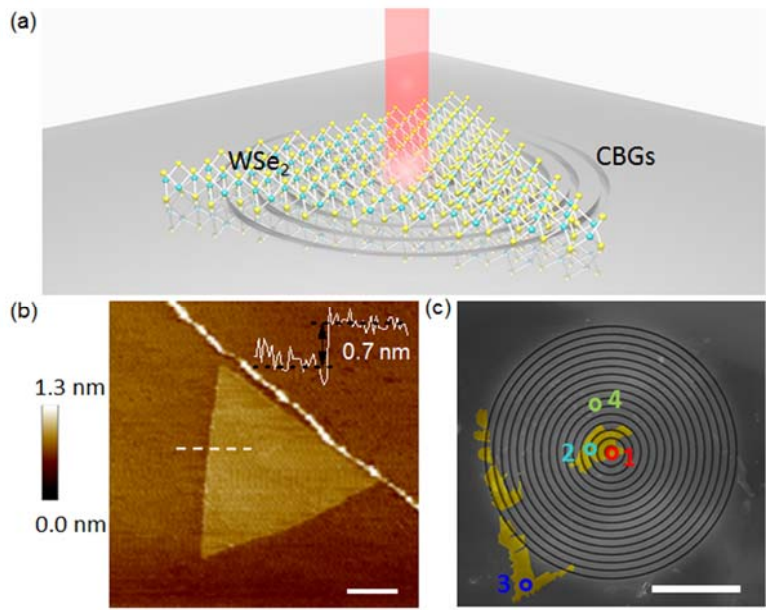

*Figure 1. a)* Schematic of a WSe$_2$ monolayer on top of a circular Bragg grating (CBG) structure. *b)* Atomic force microscope (AFM) image of the WSe$_2$ monolayer. The height profile (inset) extracted along the white dashed line shows the substrate-to-WSe$_2$ step, corresponding to a flake thickness of ~0.7 nm. *c)* False-colour scanning electron microscope (SEM) image of the WSe$_2$-CBG structure with specified positions of WSe$_2$: on the center (1), near the center (2), off the grating (3) and on the grating but away from the center (4). Scale bar in (b) and (c) is 5 μm.

A false-colour scanning electron microscopy (SEM) image of the fabricated WSe$_2$-CBG structure is shown in figure 1c. The yellow regions correspond to the WSe$_2$ monolayer, part of which covers entirely the center of the CBG (the red circle): this is essential for coupling the emission from the WSe$_2$ to the CBG photonic structure. Note that a portion of the flake also extends outside the grating structure (blue circle) which, as mentioned before, is ideal for direct comparison of the photon emission rates and directionality for the on-center and off-center (off-grating) cases. We also considered other marked areas (cyan and green circles), discussed below.

We designed the CBG to match the emission frequency of the WSe$_2$ by using a commercial software package (Lumerical Solutions, Inc.) and simulate the intrinsic optical properties of the CBG with an electric dipole source located on top of the center of the structure by 3D finite-difference time-domain (FDTD) method. The large index contrast at the $Si_3N_4$-air interface leads to strong reflections and out-of-slab-plane scattering, resulting in very high fields at the center (figure 2a). The Purcell factor is obtained from the ratio of total radiant power of a dipole in CBG to that of a dipole in vacuum by summing the transmitted power from a small box encircling the dipole. A resonant peak is clearly observed at 750 nm with the maximum Purcell factor of ~16 due to Bragg reflection and cavity resonant enhancement. Other peaks at longer wavelengths are oscillations near the Bragg reflection band-edge (figure 2b).

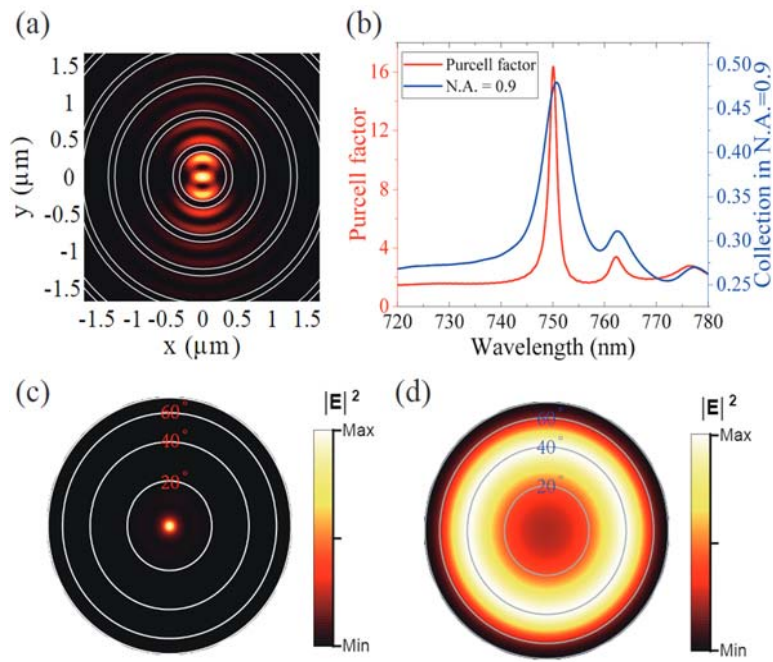

*Figure 2. a) Electric field intensity in the XY-plane superimposed to the CBG structure outline. b) Purcell factor (left axis, red) and collection efficiency (right axis, blue) calculated using an objective with NA = 0.9 for a dipole located in the center of the structure. The peak at ~750 nm is a cavity resonant mode and the others at longer wavelength are oscillations near the Bragg reflection band-edge. c, d) Far-field polar plots at the cavity resonant wavelength, 750 nm, on (c) and off (d) the CBG.*

The collection efficiency is calculated by the ratio of radiant power collected by the objective to the total power. The numerical aperture (NA) of our objective is 0.9 (corresponding to 64 degree azimuth angle in the far-field). As shown in figure 2b, a collection efficiency up to 0.5 can be achieved near the resonant wavelength as the light propagating in the $Si_3N_4$ slab is efficiently scattered out of plane within a relatively small divergence angle by the CBG. Figures 2c and 2d display the far-field patterns of a dipole emitting on and off the CBG structure, respectively. Most of the upward emitted photons are confined within an azimuthal angle of 20 degrees for the dipole on the CBG structure. On the contrary, the dipole off the center of CBG structure (indicated as off-grating) shows a donut-shape, far-field emission with a divergent emission angle wider than 60 degrees.

To evaluate the coupling of the $WSe_2$ emission to the CBG, optical characterizations of the device were performed with a lab-built scanning confocal microscope (see methods).[27] In brief, a 532-nm, continuous-wave laser is used to excite the $WSe_2$ using an air-objective with high numerical aperture (NA = 0.9) and 100x magnification. The photoluminescence (PL) signal is collected via the same objective and filtered by a 568-nm long-pass filter (to remove the residual excitation laser) and is coupled into a spectrometer through a multimode optical fibre. Optimizations for spectral overlapping were explored to maximize coupling and enhance emission. A set of CBGs was fabricated with modes ranging from 728 to 755 nm, and the resonant wavelength was examined, as show in Figure S1. We chose a specific CBG that matches the resonance of the $WSe_2$ emission, as shown in Figure 3a. The resonant mode of the CBG is visible at ~756 nm, consistent with the simulation results discussed above. The measured quality factor is $Q$ ~120, defined as $Q = \lambda/\Delta\lambda$, where $\lambda$ is the cavity resonance wavelength and $\Delta\lambda$ is the full width at half maximum (FWHM) of the resonance peak. Figure 3b shows the PL emission from the hybrid $WSe_2$-CBG device measured in ambient conditions. We investigate the PL enhancement by measuring PL of $WSe_2$ in the center (marked as position 1) of the CBG compared with that from $WSe_2$ off the center of the CBG (marked as position 3) and used as the reference value (see figure 1c). The emission from the $WSe_2$ area coupled to the cavity (red curve) is ~4.5 times higher than that of the pristine uncoupled $WSe_2$ (blue curve). The circles in figure 1c indicate where the on- (red circle, position 1) and off-grating (blue circle, position 3) PL measurements were carried out. Notably, the PL spectrum from the $WSe_2$ emission at position 1 is different from that at position 3, due to coupling of the $WSe_2$ PL emission to the CBG resonant mode.

The PL emission from the $WSe_2$ flake is temperature-dependent[28-29] and has been shown to be higher at lower temperatures[30] owing to the suppression of the non-radiative decay components and the reduction in exposure to $O_2$ and $H_2O$—all resulting in lower radiative emission from $WSe_2$.[20] Thus, we perform PL measurement on the $WSe_2$-CBG in vacuum, at low temperature (77 K) to study the resulting coupling between the $WSe_2$ and the CBG.

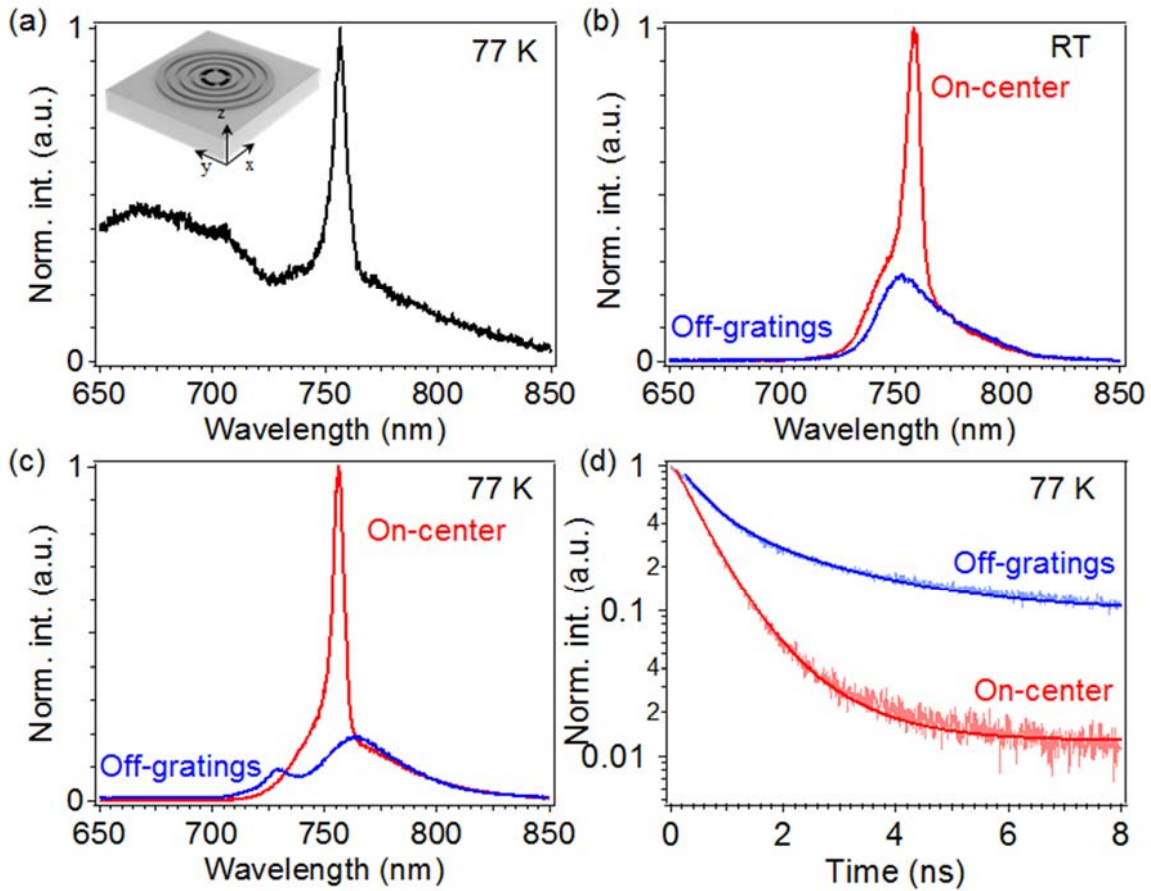

***Figure 3.*** *Optical characterization of the WSe$_2$-CBG hybrid structure. **a)** CBG mode measured from a bare CBG at 77 K with 532-nm excitation. Inset: schematic of the CBG structure with a black circle showing the area where the resonance spectrum is collected from. **b, c)** PL emission comparison for WSe$_2$ on-center (red line) and off-grating (blue line) measured at room temperature (b) and at 77 K (c). **d)** Time-resolved PL measurement from WSe$_2$ on-center (red circle) and off-grating (blue circle) at 77 K.*

The PL emission from the WSe$_2$ on-center and off-grating is shown in figure 3c. A higher, seven-fold enhancement is observed at 77 K from the on-center, coupled WSe$_2$. To further characterize the nature of the enhanced emission, time-resolved PL measurements were carried out for the on-center and off-grating WSe$_2$ using a 532-nm, pulsed laser excitation (repetition rate 40 MHz, pulse width 4 ps) at 77 K. The measured curves are shown in figure 3d both for the on-center (red) and off-grating (blue) case, confirming the reduction in the WSe$_2$ lifetime after coupling with the CBG's resonant mode. The curves are fitted with a double-exponential function, where the first exponential is fixed and determined directly by measuring the instrument response of our setup. The reduction in lifetime of the $WSe_2$ is 4 times (i.e. Purcell enhancement of ~4), approximately 1.8 times lower than the overall emission enhancement. The difference is attributed to the increased absorption of light in the CBG, as well as a directional emission from the CBG that together contribute to the increased emission intensity. As expected, the room-temperature lifetime reduction is lower (see figure. S2), resulting also in a lower overall enhancement.

To further understand the temperature-dependent excitonic properties of the $WSe_2$ coupled to the CBG, we systematically analysed temperature-resolved PL spectra of the $WSe_2$-CBG hybrid structure. Figure 4a and 4b show the PL spectra collected from the $WSe_2$ on and off the CBG center at temperatures ranging from 77 to 300 K. For the uncoupled $WSe_2$ monolayer, we observed spectral splitting at a temperature below 120 K (figure 4b), as reported elsewhere.[28-29, 31] The higher energy peak is the neutral exciton $X^0$ with a typical PL emission at ~729 nm, whereas the lower-energy peak is the trion peak, labelled X*, and emitting at ~763 nm. The trion X* emission peak is rather broad, it displays an exponential low-energy tail and it is thermally quenched for temperatures above 120 K. These features observed for the X* peak are usually considered evidence for disorder-related effects.[30]

At 77 K, the trion peak dominates the PL emission spectrum (figure 4a). The emission matches spectrally the resonant mode of the CBG at ~756 nm, hence resulting in the observed seven-fold enhancement in PL intensity. However, the trion's emission intensity steadily decreases and red-shifts from ~765 nm to ~783 nm at temperatures above 77 K (figure 4b) due to thermal activation or recombination into free exciton. We thus attribute the higher enhancement at 77 K to a combination of effects: the aforementioned increased $WSe_2$ PL and the better spectral matching between the $WSe_2$ trion and the CBG resonant mode. Figure 4c shows the overall enhancement at the resonant wavelength from the coupled and the uncoupled case as a function of temperature. As explained above, the enhancement at 77 K is much stronger, due to preferential spectral match of the cavity mode and the trion emission.

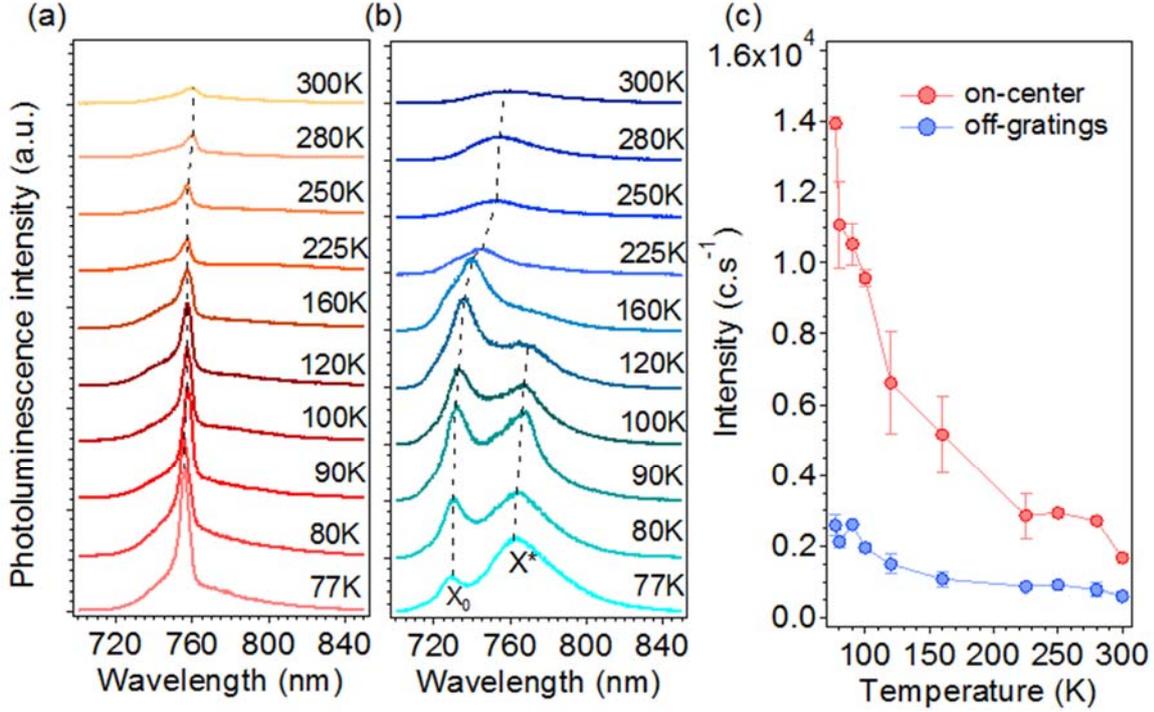

*Figure 4. Temperature-dependent emission characterization of the WSe$_2$-CBG hybrid system. a) PL emission of WSe$_2$ on the center of the CBG structure and b) off-grating measured from 300 K to 77 K. The neutral exciton ($X^0$) and the trion peak ($X^*$) are labelled. c) Temperature-dependent PL intensity at resonant wavelength of the CBG, from WSe$_2$ on-center (red circles) and off-gratings (blue circles).*

Finally, we measure the optical spin valley contrast at room temperature. The measurement setup is described in Figure S3. The degree of PL polarization with the helicity parameter is defined as

$$\rho = \frac{I(\sigma+) - I(\sigma-)}{I(\sigma+) + I(\sigma-)}$$

where $I(\sigma+)$ and $I(\sigma-)$ are the right and left circularly polarized emissions, respectively. There is only a small ($\rho$ ~3.6%) contrast between the $\sigma+$ and $\sigma-$ excitation for the uncoupled WSe$_2$, as reported in previous studies[6] (figure 5a). Note that our measurements have an estimated error of ≤ 10% due to the non-perfect circular polarization of the beam (e.g. due to reflections which slightly distort the polarization). Figure 5b shows the contrast collected from position 4 which gives a value for the contrast still within the measurement error range, $\rho$ ~8%. When the excitation is moved to position 2, ~1 μm away from the CBG center, the contrast increased to ~17% (figure 5c) possibly due to strain or partial coupling.[32] The polarization contrast then reaches an extremely high value $\rho$ ~43% at the center of the CBG (figure 5d) when the PL is totally coupled. To the best of our knowledge, this is the largest valley contrast that has been measured for WSe$_2$.

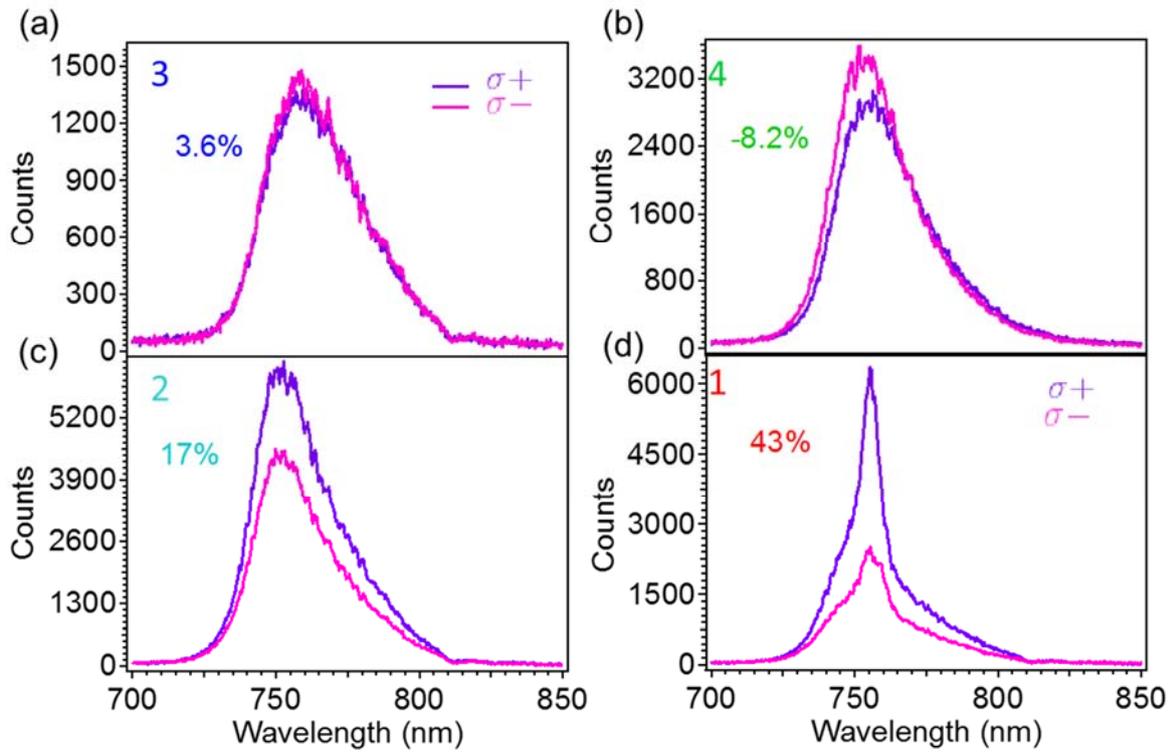

*Figure 5. a–d) Emission from WSe$_2$ at different spots (same ones of figure 1c) under circularly-polarized excitation at 532 nm. The σ+ and σ− PL components are measured at each of the spots with the corresponding values for the contrast ρ listed on each graph.*

The significantly-enhanced contrast may be explained by the way CBG structures affect polarization. Previous works studying reflection[33] and cathodoluminescence (CL)[34] from CBGs have shown that an ideal CBG excited exactly in the center has a fairly simple symmetrical output in terms of polarization distribution. The picture, however, changes dramatically if the excitation is placed even slightly off-center. CL can be a good qualitative predictor of PL, and it shows that moving the excitation off-center within the first CBG ring leads to a very complex distribution of left-hand-side and right-hand-side circular polarizations with respect to emission angles.[34] Taking into account the limited collection angle, this effect may be responsible for a strong enhancement of the contrast between the σ+ and σ− emission, when the WSe$_2$ is excited slightly off-center of the CBG. This effect is also expected to diminish when moving several rings away from the CBG center, as the CBG modes are no longer efficiently excited, and the constructive interference from multiple CBG rings diminishes. This is what we see in the experiment, as the contrast diminishes with the increasing distance from the CBG center.

WSe₂ also has a range of mechanisms such as excitation of trions[35] and up-conversion PL[36] that can lead to a drastic $\sigma+/\sigma-$ contrast increase. Although the wavelengths in these other reports differ from those in our experiments, homologous effects may play a role for the increase in polarization contrast near the center of the CBG—for instance if WSe₂ modes with strong polarization dependence couple more or less efficiently to the CBG modes. In our temperature-resolved experiments, we see a preferential coupling of the trion emission to the CBG mode—based on wavelength-matching arguments. Regardless of the exact mechanism leading to the contrast increase, our results indicate that CBGs can become a useful platform for the future development of WSe₂-based valleytronics.[37]

To conclude, we demonstrated enhanced emission from a WSe₂ monolayer integrated with a circular Bragg grating structure. Approximately a three-fold photoluminescence enhancement at room temperature and a seven-fold enhancement at 77 K are observed. Our temperature-dependent emission measurements from WSe₂ shed light onto the mechanism for such enhancement in the WSe₂-CBG hybrid system. Finally, we show that an optical readout of the spin valley is possible for the coupled WSe₂-CBG system, with a record contrast of ~40%. Our results contribute to possible, further developments of 2D materials and optical cavities for integrated nanophotonic and valleytronic applications.

## Methods
### Numerical modelling
Numerical calculations are performed using a commercial solver, Lumerical software. Refractive indices of $Si_3N_4$ and $SiO_2$ are set to 2.02 and 1.45, respectively. The parameters for silicon are taken from the software's build-in material database, Si (Silicon)–Palik. The structure comprises of a 180-nm $Si_3N_4$ CBGs, a 750-nm $SiO_2$ and a Si substrate. The CBG structure consists of a central disk with a radius of 325 nm and 16 concentric circular rings featuring a period of 450 nm and a trench of 90 nm. The simulation volume is 16.85 x 16.85 x 1.76 $\mu m^3$ with a non-uniform mesh of mesh accuracy 4. A dipole is placed 0.5 nm above the CBG as an exciting source with wavelengths ranging from 720 nm to 780 nm. A field monitor at the center of $Si_3N_4$ CBGs is set to record the near field and a power monitor 100 nm above the CBG is used to record up-going near field for performing near- to far-field projections. A transmission box consisting of 6 power monitors in 6 directions are used for calculating the total radiant power. The collection efficiency calculated based on a NA = 0.9 objective is obtained from the ratio of up-going power in 64 degree azimuthal angle to the total emitted power.

**Design and fabrication of CBGs**

750-nm $SiO_2$ and 180-nm $Si_3N_4$ are deposited on a silicon wafer at 300 in an inductively-coupled plasma chemical vapour deposition. The patterns of the CBGs are defined by E-beam lithography with a positive tone resist (ZEP 520A) in a 100-KeV writer (Vistec EBPG 5000+) and transferred into the $Si_3N_4$ layer using a $CHF_3$-$O_2$ based try etch process. Finally, the e-beam resist is removed by a 1165 remover solution in an ultrasonication bath.

**Optical characterization**

The optical characterizations of the structures were carried out with a lab-built scanning confocal microscope. A continuous wave (CW) 532-nm laser (Gem 532, Laser Quantum Ltd.) was used for excitation. The laser was directed through a Glan-Taylor polarizer (Model No. DGL10, Thorlabs Inc.) and a half-wave plate, and focused onto the sample using a high-numerical-aperture objective lens (NA = 0.9, TU Plan Flour 100x, Nikon). The laser was scanned across the sample using an x−y piezo scanning mirror (FSM-300, Newport). The collected light was filtered using a 532-nm dichroic mirror (532-nm laser BrightLine, Semrock) and an additional 568-nm long pass filter (LP Filter 568 nm, Semrock). The signal was then coupled into a graded-index fiber, where the fiber aperture served as a pinhole. A fiber splitter was used to direct the light to a spectrometer (Acton SpectraPro, Princeton Instrument Inc.) and to an avalanche photodiodes (SPCM-AQRH-14-FC, Excelitas Technologies) for time-resolved measurement. To avoid sample damage, the laser excitation was performed at low power (<50 μW). Lifetime measurements were performed using a Fianium Whitelase Supercontinuum laser (NKT Photonics) coupled into a fibre. The laser was collimated and went through a short-pass filter (568 nm, Samrock) to filter the 532-nm light. The repetition rate was kept at 20 MHz for all the experiments. Samples for radiative lifetime measurement were all measured at 77 K in vacuum. The data were fit by using a double exponential function. Temperature-dependant experiments were also performed using a closed-cycle refrigerating system cryostat cooled with liquid nitrogen.

**Acknowledgements**

The authors thanks financial support from the Australian Research council (*via* DP180100077, DE180100070, DE 1801100810) the Asian Office of Aerospace Research and Development grant FA2386-17-1-4064, the Office of Naval Research Global under grant number N62909-18-1-2025 are gratefully acknowledged. This research is supported by an Australian Government Research Training Program Scholarship. The authors thank Dr. Toan Trong Tran for fruitful discussions and the schematic in figure 1a.

**References**

1. Fiori, G.; Bonaccorso, F.; Iannaccone, G.; Palacios, T.; Neumaier, D.; Seabaugh, A.; Banerjee, S. K.; Colombo, L., Electronics based on two-dimensional materials. *Nature nanotechnology* **2014,** *9* (10), 768-79.
2. Zhang, Y.; Chang, T. R.; Zhou, B.; Cui, Y. T.; Yan, H.; Liu, Z.; Schmitt, F.; Lee, J.; Moore, R.; Chen, Y.; Lin, H.; Jeng, H. T.; Mo, S. K.; Hussain, Z.; Bansil, A.; Shen, Z. X., Direct observation of the transition from indirect to direct bandgap in atomically thin epitaxial MoSe2. *Nature nanotechnology* **2014,** *9* (2), 111-5.
3. He, K.; Kumar, N.; Zhao, L.; Wang, Z.; Mak, K. F.; Zhao, H.; Shan, J., Tightly bound excitons in monolayer WSe(2). *Physical review letters* **2014,** *113* (2), 026803.
4. Kormányos, A.; Zólyomi, V.; Drummond, N. D.; Rakyta, P.; Burkard, G.; Fal'ko, V. I., Monolayer MoS2: Trigonal warping, theΓvalley, and spin-orbit coupling effects. *Physical Review B* **2013,** *88* (4), 045416.
5. Jones, A. M.; Yu, H.; Ghimire, N. J.; Wu, S.; Aivazian, G.; Ross, J. S.; Zhao, B.; Yan, J.; Mandrus, D. G.; Xiao, D.; Yao, W.; Xu, X., Optical generation of excitonic valley coherence in monolayer WSe2. *Nature nanotechnology* **2013,** *8* (9), 634-8.
6. Zeng, H.; Dai, J.; Yao, W.; Xiao, D.; Cui, X., Valley polarization in MoS2 monolayers by optical pumping. *Nature nanotechnology* **2012,** *7* (8), 490-3.
7. Yuan, H.; Bahramy, M. S.; Morimoto, K.; Wu, S.; Nomura, K.; Yang, B.-J.; Shimotani, H.; Suzuki, R.; Toh, M.; Kloc, C.; Xu, X.; Arita, R.; Nagaosa, N.; Iwasa, Y., Zeeman-type spin splitting controlled by an electric field. *Nature Phys.* **2013,** *9* (9), 563-569.
8. Zhang, Y. J.; Oka, T.; Suzuki, R.; Ye, J. T.; Iwasa, Y., Electrically Switchable Chiral Light-Emitting Transistor. *Science* **2014,** *344* (6185), 725-728.
9. Ross, J. S.; Wu, S.; Yu, H.; Ghimire, N. J.; Jones, A. M.; Aivazian, G.; Yan, J.; Mandrus, D. G.; Xiao, D.; Yao, W.; Xu, X., Electrical control of neutral and charged excitons in a monolayer semiconductor. *Nature communications* **2013,** *4*, 1474.
10. Wu, S.; Ross, J. S.; Liu, G.-B.; Aivazian, G.; Jones, A.; Fei, Z.; Zhu, W.; Xiao, D.; Yao, W.; Cobden, D.; Xu, X., Electrical tuning of valley magnetic moment through symmetry control in bilayer MoS2. *Nature Physics* **2013,** *9* (3), 149-153.
11. Morpurgo, A. F., Gate control of spin-valley coupling. *Nature Physics* **2013,** *9* (9), 532-533.
12. Wang, Z.; Shan, J.; Mak, K. F., Valley- and spin-polarized Landau levels in monolayer WSe2. *Nature nanotechnology* **2016,** *12*, 144.
13. Schaibley, J. R.; Yu, H.; Clark, G.; Rivera, P.; Ross, J. S.; Seyler, K. L.; Yao, W.; Xu, X., Valleytronics in 2D materials. *Nature Reviews Materials* **2016,** *1*, 16055.
14. Schaibley, J. R.; Rivera, P.; Yu, H.; Seyler, K. L.; Yan, J.; Mandrus, D. G.; Taniguchi, T.; Watanabe, K.; Yao, W.; Xu, X., Directional interlayer spin-valley transfer in two-dimensional heterostructures. *Nature communications* **2016,** *7*, 13747.
15. Wu, S.; Buckley, S.; Schaibley, J.; Feng, L.; Yan, J. Q.; G Mandrus, D.; Hatami, F.; Yao, W.; Vuckovic, J.; Majumdar, A.; Xu, X., *Monolayer semiconductor nanocavity lasers with ultralow thresholds*. 2015; Vol. 520.
16. Ross, J. S.; Klement, P.; Jones, A. M.; Ghimire, N. J.; Yan, J.; Mandrus, D. G.; Taniguchi, T.; Watanabe, K.; Kitamura, K.; Yao, W.; Cobden, D. H.; Xu, X., Electrically tunable excitonic light-emitting diodes based on monolayer WSe2 *p-n* junctions. *Nat Nanotechnol* **2014,** *9* (4), 268-72.

17. Xu, Z.-Q.; Zhang, Y.; Wang, Z.; Shen, Y.; Huang, W.; Xia, X.; Yu, W.; Xue, Y.; Sun, L.; Zheng, C.; Lu, Y.; Liao, L.; Bao, Q., Atomically thin lateral p–n junction photodetector with large effective detection area. *2D Materials* **2016**, *3* (4), 041001.
18. Sanfeng, W.; Sonia, B.; Aaron, M. J.; Jason, S. R.; Nirmal, J. G.; Jiaqiang, Y.; David, G. M.; Wang, Y.; Fariba, H.; Jelena, V.; Arka, M.; Xiaodong, X., Control of two-dimensional excitonic light emission via photonic crystal. *2D Materials* **2014**, *1* (1), 011001.
19. Chen, H.; Nanz, S.; Abass, A.; Yan, J.; Gao, T.; Choi, D.-Y.; Kivshar, Y. S.; Rockstuhl, C.; Neshev, D. N., Enhanced Directional Emission from Monolayer WSe2 Integrated onto a Multiresonant Silicon-Based Photonic Structure. *ACS Photonics* **2017**, *4* (12), 3031-3038.
20. Butun, S.; Tongay, S.; Aydin, K., Enhanced Light Emission from Large-Area Monolayer MoS2 Using Plasmonic Nanodisc Arrays. *Nano letters* **2015**, *15* (4), 2700-2704.
21. Tran, T. T.; Choi, S.; Scott, J. A.; Xu, Z.-Q.; Zheng, C.; Seniutinas, G.; Bendavid, A.; Fuhrer, M. S.; Toth, M.; Aharonovich, I., Room-Temperature Single-Photon Emission from Oxidized Tungsten Disulfide Multilayers. *Advanced Optical Materials* **2017**, *5* (5), 1600939.
22. Su, M. Y.; Mirin, R. P., Enhanced light extraction from circular Bragg grating coupled microcavities. *Applied Physics Letters* **2006**, *89* (3), 033105.
23. Bauer, C.; Giessen, H.; Schnabel, B.; Kley, E. B.; Schmitt, C.; Scherf, U.; Mahrt, R. F., A Surface-Emitting Circular Grating Polymer Laser. *Advanced materials* **2001**, *13* (15), 1161-1164.
24. Tien, P. K., Method of forming novel curved-line gratings and their use as reflectors and resonators in integrated optics. *Optics Letters* **1977**, *1* (2), 64-66.
25. Huang, J. K.; Pu, J.; Hsu, C. L.; Chiu, M. H.; Juang, Z. Y.; Chang, Y. H.; Chang, W. H.; Iwasa, Y.; Takenobu, T.; Li, L. J., Large-area synthesis of highly crystalline WSe(2) monolayers and device applications. *ACS nano* **2014**, *8* (1), 923-30.
26. Xu, Z. Q.; Zhang, Y.; Lin, S.; Zheng, C.; Zhong, Y. L.; Xia, X.; Li, Z.; Sophia, P. J.; Fuhrer, M. S.; Cheng, Y. B.; Bao, Q., Synthesis and Transfer of Large-Area Monolayer WS2 Crystals: Moving Toward the Recyclable Use of Sapphire Substrates. *ACS nano* **2015**, *9* (6), 6178-87.
27. Tran, T. T.; Wang, D.; Xu, Z.-Q.; Yang, A.; Toth, M.; Odom, T. W.; Aharonovich, I., Deterministic Coupling of Quantum Emitters in 2D Materials to Plasmonic Nanocavity Arrays. *Nano letters* **2017**, *17* (4), 2634-2639.
28. Godde, T.; Schmidt, D.; Schmutzler, J.; Aßmann, M.; Debus, J.; Withers, F.; Alexeev, E. M.; Del Pozo-Zamudio, O.; Skrypka, O. V.; Novoselov, K. S.; Bayer, M.; Tartakovskii, A. I., Exciton and trion dynamics in atomically thin MoSe2 and WSe2: Effect of localization. *Physical Review B* **2016**, *94* (16), 165301.
29. Yan, T.; Qiao, X.; Liu, X.; Tan, P.; Zhang, X., Photoluminescence properties and exciton dynamics in monolayer WSe2. *Applied Physics Letters* **2014**, *105* (10), 101901.
30. Huang, J.; Hoang, T. B.; Mikkelsen, M. H., Probing the origin of excitonic states in monolayer WSe2. *Scientific reports* **2016**, *6*, 22414.
31. Zhao, W.; Ribeiro, R. M.; Toh, M.; Carvalho, A.; Kloc, C.; Castro Neto, A. H.; Eda, G., Origin of Indirect Optical Transitions in Few-Layer MoS2, WS2, and WSe2. *Nano Lett.* **2013**, *13* (11), 5627-5634.
32. Zhu, C. R.; Wang, G.; Liu, B. L.; Marie, X.; Qiao, X. F.; Zhang, X.; Wu, X. X.; Fan, H.; Tan, P. H.; Amand, T.; Urbaszek, B., Strain tuning of optical emission energy and polarization in monolayer and bilayer MoS${}_{2}$. *Physical Review B* **2013**, *88* (12), 121301.
33. Osorio, C. I.; Mohtashami, A.; Koenderink, A. F., K-space polarimetry of bullseye plasmon antennas. *Scientific reports* **2015**, *5*, 9966.
34. Osorio, C. I.; Coenen, T.; Brenny, B. J. M.; Polman, A.; Koenderink, A. F., Angle-Resolved Cathodoluminescence Imaging Polarimetry. *ACS Photonics* **2016**, *3* (1), 147-154.
35. Courtade, E.; Semina, M.; Manca, M.; Glazov, M. M.; Robert, C.; Cadiz, F.; Wang, G.; Taniguchi, T.; Watanabe, K.; Pierre, M.; Escoffier, W.; Ivchenko, E. L.; Renucci, P.; Marie, X.; Amand, T.; Urbaszek, B., Charged excitons in monolayer WSe2: Experiment and theory. *Physical Review B* **2017**, *96* (8), 085302.
36. Manca, M.; Glazov, M. M.; Robert, C.; Cadiz, F.; Taniguchi, T.; Watanabe, K.; Courtade, E.; Amand, T.; Renucci, P.; Marie, X.; Wang, G.; Urbaszek, B., Enabling valley selective exciton scattering in monolayer WSe2 through upconversion. *Nature communications* **2017**, *8*, 14927.

37. McCreary, K. M.; Currie, M.; Hanbicki, A. T.; Chuang, H.-J.; Jonker, B. T., Understanding Variations in Circularly Polarized Photoluminescence in Monolayer Transition Metal Dichalcogenides. *ACS nano* **2017,** *11* (8), 7988-7994.